\documentclass[twocolumn,floatfix,superscriptaddress,amsmath,amssymb,aps,prbrc]{revtex4-1}
\usepackage{psfrag}
\usepackage{wrapfig}
\usepackage{subfigure}
\usepackage{psfrag}
\usepackage{tabularx}
\usepackage[dvips]{graphicx}
\DeclareGraphicsExtensions{.eps}
\def\LSCO{La$_{2-x}$Sr$_x$CuO$_4$}
\def\LBCO{La$_{2-x}$Ba$_x$CuO$_4$}

\def\C60{A$_x$C$_{60}$}
\def\LNSCO{La$_{1.6-x}$Nd$_{0.4}$Sr$_x$CuO$_{4}$}

\def\LNSCO{La$_{1.6-x}$Nd$_{0.4}$Sr$_x$CuO$_{4}$}

\def\HgCu3{HgCa$_2$Cu$_3$O$_{8+y}$}
\def\HgCu4{HgBa$_2$Ca$_3$Cu$_4$O$_{10+y}$}
\def\TlCu{Tl$_2$Ba$_2$CuO$_{6+\delta}$}
\def\TlCu3{Tl$_2$Ba$_2$Ca$_2$Cu$_3$O$_{10+y}$}
\def\TlCu4{Tl$_2$Ba$_2$Ca$_3$Cu$_4$O$_{12+y}$}

\def\BiCu3{Bi$_2$Sr$_2$Ca$_{2}$Cu$_3$O$_y$}

\def\8LSCO{La$_{1.88}$Sr$_{.12}$CuO$_4$}
\def\110LNSCO{La$_{1.5}$Nd$_{0.4}$Sr$_{0.1}$CuO$_{4}$}
\def\stage4LCO{La$_{2}$CuO$_{4+\delta}$}
\def\Y248{YBa$_2$Cu$_4$O$_8$}

\def\NbSe2{NbSe$_2$}
\def\TaSe2{TaSe$_2$}
\def\TiSe2{TiSe$_2$}
\def\NaCoOH2O{Na$_{0.3}$CoO$_{2y}$H$_2$O}
\def\MgB2{MgB${}_2$}

\begin{document}
\title{
The role of nematic fluctuations in the thermal melting of pair-density-wave phases in two-dimensional superconductors}
\author{Daniel G.\ Barci}
\affiliation{Departamento de F{\'\i}sica Te\'orica,
Universidade do Estado do Rio de Janeiro, Rua S\~ao Francisco Xavier 524, 20550-013,  Rio de Janeiro, RJ, Brazil.}
\author{Eduardo Fradkin}
\affiliation{Department of Physics, University of Illinois 
at Urbana-Champaign, 1110 W. Green Street, Urbana, Illinois 61801-3080, U.S.A.}
\date{\today}
\begin{abstract}
We study properties of phase transitions of 2D superconductor liquid crystal phases, and analyze the competition between the recently proposed  Pair Density Wave (PDW)  and  nematic $4e$ superconductor ($4e$SC). 
Nematic fluctuations enhance the $4e$SC and suppress the PDW phase. 
In the absence of lattice effects, the PDW state exists only at $T=0$ and the low temperature phase is a nematic $4e$SC with short ranged PDW order. 
A geometric description of the $4e$ SC is presented.
\end{abstract}
\maketitle

A beautiful series of experiments\cite{li-2007,tranquada-2008} have shown that the cuprate superconductor {\LBCO} exhibits a remarkable dynamical 
layer decoupling behavior 
near the $x=1/8$ ``anomaly''. In this regime 
the onset of static charge and spin stripe 
order coincides with the development of  an extreme transport anisotropy. 
Similar effects have been seen in stripe-ordered {\LNSCO} \cite{tajima-2001,ding-2008} and
in the magnetic-field induced stripe-ordered phase of {\LSCO} \cite{schafgans-2008}.

The remarkable layer-decoupling effect suggests that the inter-layer Josephson coupling is somehow frustrated when SC order is forced to coexist with 
charge and/or spin stripe order. It was proposed by Berg {\it et al} \cite{berg-2007,berg-2008a,berg-2009b}, that this effect can 
be understood naturally by postulating that the SC order also becomes ``striped'' and that all three orders rather than competing are intertwined. 
In the resulting striped SC 
state, a pair-density-wave (PDW), the (unidirectional) SC pair field oscillates in space with an ordering wave vector $\mathbf{Q}$.
The PDW has the same structure as  an ``Fulde, Ferrell, Larkin, Ovchinnikov''  state, (FFLO)\cite{fulde-1964,larkin-1964} except that in this case the spatial 
modulation of the 
Cooper pairing is not due to Fermi surfaces mismatched by a (Zeeman) magnetic field (as in the standard FFLO scenario). Instead, the PDW arises in a strong 
coupling regime 
where the BCS mechanism is not 
effective: the modulation of the SC state
is due to the same physics behind the formation of inhomogeneous states in doped 
Mott insulators\cite{emery-1993,kivelson-1998}. 
 FFLO states
have been proposed to occur in imbalanced  cold atom Fermi systems with different species\cite{radzihovsky-2008} and in heavy fermions systems when 
different orbitals hybridize under external pressure\cite{padilha-2009}. In Ref. \onlinecite{berg-2009}, the phase diagram of this problem was studied 
deep in the (Ising) nematic phase
assuming that the coupling to the lattice is so strong 
that it completely suppresses  the fluctuations of nematic (orientational) order. 

In this letter, we 
consider the role of nematic fluctuations on the structure of the phase diagram associated with the PDW state, and on the properties of its phases. 
Here we focus on the interplay between nematic, PDW and the 
charge 4e SC which we will show also to have nematic character.  
 We consider different regimes characterized by the strength of the coupling between the orientational (nematic) degrees of freedom of the 
 PDW with the underlying lattice, all the way to the 
decoupled case where the system has a continuous rotational invariance. 
In the absence of a coupling to a lattice it is not possible to break spontaneously translation invariance in 2D, and a 2D continuum smectic  
is always thermally melted by  
proliferation of (finite energy) dislocations\cite{nelson-1981}. Thus, the topology of the phase diagram of the PDW state is strongly affected by the coupling 
to the lattice.  

The PDW is an anisotropic quantum liquid  crystal state that breaks the point group symmetry of the lattice as well as
translation and global gauge invariance. We have constructed a phenomenological theory of the PDW with a structure reminiscent of the 
McMillan-de Gennes theory of  the nematic/smectic  
transition of classical liquid crystals, with several significant differences: a) it involves the order parameter of the charge $4e$ SC and 
its geometric coupling to nematic order, b) the nematic order has, in addition to the standard elastic Frank free energy\cite{degennes-1993}, 
a coupling that breaks the continuous rotational invariance down to the (Ising) point group symmetry of the lattice.
The strength $h$ of the lattice coupling defines a temperature $T_h$, below which, the coupling to the lattice  
breaks the continuous rotational invariance down to the point group symmetry (say $\mathbb{Z}_2$) of the lattice, 
forcing the director to 
take only one of two perpendicular orientations.
In electronic nematic phases, such as those occurring in strongly correlated systems, lattice effects are not normally small. Typically
  nematic order occurs on {\em mesoscopic scales}, and the {\em effective} value of $h$  can be smaller than the other couplings.
As temperature increases, Ising domain walls proliferate and, above $T_h$ where
the domain wall tension vanishes, the Ising order disappears
and the system becomes invariant under $90^\circ$ rotations. We discuss the behavior and role of the topological excitations of this system as $h$ is varied 
from $0$ to large values.
\begin{figure}[hbt]
\psfrag{TH}{$\frac{T_h}{\rho_s}$}
\psfrag{K}{$\frac{\kappa}{\rho_s}$}
\psfrag{T}{$\frac{T}{\rho_s}$}
\psfrag{NIS}{\footnotesize Normal Isotropic}
\psfrag{P}{\footnotesize $P$}
\psfrag{M}{\footnotesize $M$}
\psfrag{M'}{\footnotesize $M'$}
\psfrag{PDW}{\footnotesize PDW}
\psfrag{N-SC}{\footnotesize N-SC}
\psfrag{SC}{\footnotesize SC}
\psfrag{Nematic}{\footnotesize Nematic}
\psfrag{Isotropic}{\footnotesize Isotropic}
\psfrag{CDW}{\footnotesize CDW}
\begin{center}
\includegraphics[width=0.3\textwidth]{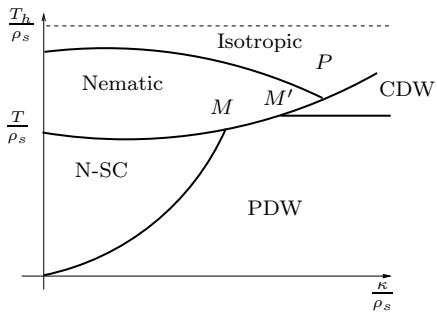}
\end{center}
\caption{Schematic 2D phase diagrams for high $T_h$:  PDW/N-SC, Nematic, CDW are KT transitions. The Isotropic-Nematic line is an Ising transition and the 
 Isotropic-CDW is first order.}
 \label{fig1}
 \end{figure}
\begin{figure}[hbt]
\psfrag{TH}{$\frac{T_h}{\rho_s}$}
\psfrag{K}{$\frac{\kappa}{\rho_s}$}
\psfrag{T}{$\frac{T}{\rho_s}$}
\psfrag{NIS}{\footnotesize Normal Isotropic}
\psfrag{P}{\footnotesize $P$}
\psfrag{M}{\footnotesize $M$}
\psfrag{M'}{\footnotesize $M'$}
\psfrag{PDW}{\footnotesize PDW}
\psfrag{N-SC}{\footnotesize N-SC}
\psfrag{SC}{\footnotesize SC}
\psfrag{Nematic}{\footnotesize Nematic}
\psfrag{Isotropic}{\footnotesize Isotropic}
\psfrag{CDW}{\footnotesize CDW}
\begin{center}
\includegraphics[width=0.3\textwidth]{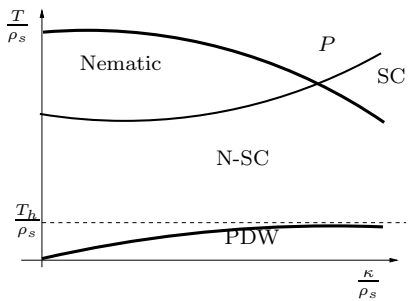}
\end{center}
 \caption{Schematic 2D phase diagram for  low $T_h$. Labels are the same as in Fig.\ref{fig1}.}
 \label{fig2}
\end{figure}

\begin{figure}[hbt]
\psfrag{TH}{$\frac{T_h}{\rho_s}$}
\psfrag{K}{$\frac{\kappa}{\rho_s}$}
\psfrag{T}{$\frac{T}{\rho_s}$}
\psfrag{NIS}{\footnotesize Normal Isotropic}
\psfrag{P}{\footnotesize $P$}
\psfrag{M}{\footnotesize $M$}
\psfrag{M'}{\footnotesize $M'$}
\psfrag{PDW}{\footnotesize PDW}
\psfrag{N-SC}{\footnotesize N-SC}
\psfrag{SC}{\footnotesize SC}
\psfrag{Nematic}{\footnotesize Nematic}
\psfrag{Isotropic}{\footnotesize Isotropic}
\psfrag{CDW}{\footnotesize CDW}
\begin{center}
\includegraphics[width=0.3\textwidth]{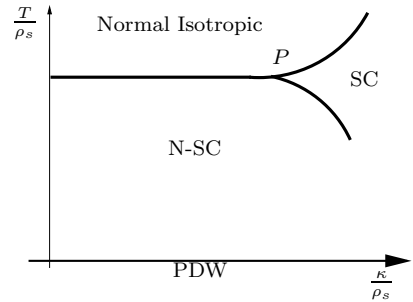}
\end{center}
 \caption{Schematic 2D phase diagrams for $T_h=0$: all lines are KT transitions. Labels are the same as in Fig.\ref{fig1}.} 
 \label{fig3}
\end{figure}

The qualitative structure of the phase diagrams is summarized in Figs.\ref{fig1}, \ref{fig2}, and \ref{fig3}.  
In the regime in which $T_h$ is large (Fig.\ref{fig1}a), the nematic is the ordered phase with higher $T_c$. 
In this regime the physics is qualitatively similar to that of Ref.\cite{berg-2009}. 
The main effect of nematic fluctuations is to soften the  smectic stiffness ($\kappa$) 
to $\kappa_{\rm eff}=\kappa/(1+\kappa |\textbf{Q}|^2/h)$ where  $\textbf{Q}$ is the 
the smectic ordering wave vector.
As the strength of the coupling constant $h$ is reduced,  the  effective CDW stiffness $\kappa_{\rm eff}$ gets weaker,
 the PDW portion of the phase diagram shrinks and the region of N-SC grows.   
There is also an isotropic/Ising-nematic transition that  
takes place at temperatures higher than the SC transition.
For large $\kappa_{\rm eff}$, as the temperature $T$ increases the PDW phase may melt  (or may not) by a direct PDW to isotropic transition or through an intermediate CDW depending on the sign of the coupling between the order parameters in the Landau theory. The 
PDW to  nematic superconductor (N-SC), 
CDW and  nematic, are Kosterlitz-Thouless (KT) transitions, mediated by unbinding of double dislocations, vortices or half-vortices-single dislocations\cite{berg-2009,radzihovsky-2008,agterberg-2008} respectively. The direct CDW/isotropic transition is likely first order. 

For lower values of $T_h$  we have a softer nematic (Fig.\ref{fig2}). Still, $T_h$ is an upper bound for  the $T_c$ of the PDW since, above this temperature, no translational order is possible. Similarly, the CDW phase is also absent.

In the extreme case of a system decoupled from the underlying lattice, 
$T_h=0$, the otherwise logarithmically divergent dislocations have now a finite energy and proliferate at all 
temperatures. In this case, the PDW phase occurs only at $T=0$ as shown in Fig. \ref{fig3}. Provided  
$\rho_s$ remains finite, in this regime  the non-SC nematic phase also cannot exist since, as we will see below, in the N-SC vortices are strongly bound to disclinations, and thus  proliferate simultaneously. Hence, at $T_h=0$ the nematic phase disappears. 
In addition, a tetracritical point is now possible: nematic and SC order are decoupled and the N-SC is a coexistence phase. 

\paragraph{Order parameters and Landau theory:} The PDW state has two complex order parameters
$\Delta_{\pm \mathbf{Q}}(r)\sim |\Delta| \exp(i\theta_{\pm \mathbf{Q}} (\mathbf{r}) )$, with $\mathbf{Q}=Q \mathbf{n}$ and $\mathbf{n}$ is the director of the nematic order. 
We now define the SC phase field
$\theta=(\theta_{+Q}+\theta_{-Q})/2$ and the smectic phase field $\varphi=(\theta_{+Q}-\theta_{-Q})/2$ \cite{berg-2009,radzihovsky-2008,agterberg-2008}. 
In this work, unlike Ref. \cite{berg-2009},  the director field $\mathbf{n}=(\cos\alpha,\sin{\alpha})$ is allowed to fluctuate.
The coupling of the orientational degree of freedom $\hat n$ couples to the lattice through a potential $V(\mathbf{n})= -(h/8)\cos(4\alpha)$ with $h>0$, that reduces the continuous rotational symmetry to a discrete subgroup, i.e. 
$\mathbb{Z}_2$ for a square lattice.
For $h \neq 0$ the Ising nematic has a critical temperature $T_h>0$, 
above which the Ising degrees of freedom disorder \cite{granato-1991,garel-1980}. 
The structure
of the phase diagram depends on the values of $T_h$, the superconductor stiffness $\rho_s$ and the CDW stiffness $\kappa$.

Aside from standard quadratic and quartic terms in the PDW order parameters $\Delta_{\pm \mathbf{Q}}$, the CDW order parameter $\rho_{\mathbf{K}}$, and the charge $4e$ N-SC order parameter $\Delta_{4e}$, the PDW free energy contain trilinear couplings that relate all the order parameters leading, in particular, to the relation $\mathbf{K}=2\mathbf{Q}$, and 
the PDW order induces a uniform charge $4e$ N-SC\cite{berg-2009b}. The phase fluctuations of the CDW and the $4e$SC are locked to $2\varphi$ and $2\theta$ respectively. 
In addition, we 
consider a nematic order parameter, 
the symmetric traceless tensor $\hat N_{ij}=N(\hat n_i\hat n_j-\delta_{ij}/2)$ where $i,j=1,2$ for $(\hat x,\hat y)$ orientations respectively. 
For $T<< T_h$, $N_{ij}$ has essentially  two values $\pm N$ characterizing an Ising symmetry. 
The competition between the nematic and the $4e$ SC is governed by the quartic term $\beta {\rm Tr}N^2 |\Delta_{4e}|^2$. If $\beta<0$, the condensation of one phase enhances the other.
For $T_h> T_c^{SC}$ a multicritical point  may occur.

\paragraph{Nematic-SC couplings:} 
Local nematic order can be regarded as 
a fluctuating metric $g_{ij}=\delta_{ij}+\lambda N_{ij}$ describing the local anisotropy of the stiffness terms of the free energy
\begin{eqnarray}
F_d&=&\int dx^2\;\sqrt{{\rm det} g}\; g^{ij}\left\{ \left(D^{cdw}_i\rho_K\right)^*\left(D^{cdw}_j\rho_K\right)+ \right. \label{DC} 
\\ &&   \nonumber \\ &+&\left. \left(D^{sc}_i\Delta_{\pm Q}\right)^*\left(D^{sc}_j\Delta_{\pm Q}\right) 
+ \left(D^{4e}_i\Delta_{4e}\right)^*\left(D^{4e}_j\Delta_{4e}\right)\right\} \nonumber 
\end{eqnarray}
where the covariant derivatives are given by
\begin{eqnarray}
D_i^{sc}&=&  \nabla_i- i 2e  A_i \pm i Q\; \delta n_i   \label{Dsc}\\
D_i^{cdw}&=& \nabla_i+i 2 Q\;  \delta  n_i   \label{Dcdw} \\
D_i^{4e}&=& \nabla_i- i 4e A_i \label{D4e}
\end{eqnarray}
Here, $\mathbf{A}$ is the electromagnetic vector potential, and  $\delta \mathbf{n}=\mathbf{n}-\mathbf{n}_0$ is the  director fluctuation about an arbitrary direction $\mathbf{n}_0$. The $\pm$ sign in Eq.\eqref{Dsc} depends on whether the covariant derivative  acts on $\Delta_{\pm \mathbf{Q}}$ respectively.

The effect of nematicity is, in several aspects,  very similar with the effect that a curved surface has on an order parameter\cite{kamien-2002}: due to geometrical frustration,  disclinations can be regarded as representing an intrinsic curvature  of the geometry whose sign is the sign of its topological charge of the disclination\cite{park-1996}. From Eqs.\eqref{DC}, \eqref{Dcdw} and  \eqref{D4e}, the SC phase field couples to the nematic order through the metric, whereas the CDW order parameter also couples to the nematic in the standard way through the covariant derivative\cite{degennes-1993,turner-2008}. Thus, vortices and disclinations have an attractive (``gravitational'') interaction.

The way that $\delta \mathbf{n}$ enters in the covariant derivatives resembles the McMillan-DeGennes\cite{McMillan-1972,degennes-1993} theory for the smectic-nematic phase transition in classical liquid crystals. For small rotations, $\alpha<< 1$, $\delta \mathbf{n} \cdot \mathbf{n}_0=0+O(\alpha^2)$ and  the metric only represents a rigid anisotropy. At this level of approximation,  smectic and SC degrees of freedom are only coupled by topological constraints\cite{berg-2009}.  This limit is suitable to describe smectic-nematic phase transition. However, to study the properties of the phase, we need to consider the full structure of $\delta\mathbf{n}$.  Also, note that in Eqs. \eqref{Dcdw} and \eqref{D4e}, the ``gauge fields'' couple with the twice the value of the smectic and the electric charge. This is a direct consequence of global translation and gauge invariance that force the CDW wave vector $\mathbf{K}=2\mathbf{Q}$, and the superconductor charge to be $4e$.   
The factor ${\rm det}(g)=1-\lambda^2 N^2/4$
is important near the isotropic-nematic phase transition where $N$ fluctuates strongly.

\paragraph{PDW fluctuations:}
Deep in the PDW phase, we take the amplitudes of the order parameters $|\Delta_{\pm \mathbf{Q}}|=\Delta$,
$|\Delta_{4e}|$ and $|\rho_{\mathbf{K}}|=\rho$ as constants. The low energy physics is governed by  SC ($\theta$) and  smectic ($\varphi$) phase fluctuations. Since the temperature $T<<T_h$,  $\hat n_0$ points in the direction $\alpha=0$ or $\alpha=\pi/2$. The potential $V(\hat n_0+\delta \mathbf{n})$ induces  a ``mass'' term  $\sim h |\delta \mathbf{n}|^2$. From Eq. \eqref{DC} we obtain,  
\begin{eqnarray}
F&=&\int d^2x\;\left\{ \frac{\rho_s}{2}g^{ij}\;(\partial_i \theta-2e A_i)(\partial_j \theta-2e A_j)\right. \nonumber \\ 
&+& \frac{\kappa}{2}g^{ij}\;(\partial_i \varphi+ Q\delta n_i)(\partial_j \varphi+ Q\delta n_j)\nonumber \\
&+&\left. K_1  \left(\mathbf{\nabla} \cdot \delta\mathbf{n}\right)^2 
+ K_3 \left(\mathbf{\nabla} \times \delta\mathbf{n} \right)^2 + h|\delta \mathbf{n}|^2\right\}
\label{Fmassive}
\end{eqnarray}
where $\rho_s=\Delta^2+|\Delta_{4e}|^2$ is the superfluid stiffness of the $4e$ SC, and $\kappa=\Delta^2+\rho^2$ is the CDW stiffness; $K_1$ and $K_3$ are the nematic Frank constants\cite{degennes-1993}. In what follows we will assume that both $\rho_s>0$ and $\kappa>0$ and that they never become small.
We can now safely integrate over  the massive nematic fluctuations. Let $\mathbf{n}_0$ be in the $x$ direction. By expanding $\delta \mathbf{n}$ to leading order in $\alpha$, we 
obtain in the long wavelength limit 
\begin{eqnarray}
F&=&\int d^2x\;\left\{ \rho_s\left(\lambda_s\;(\partial_x \theta)^2 +\lambda_s^{-1}(\partial_y \theta)^2\right)\right. \nonumber \\ 
&+&\left. \kappa_{\rm eff}\left( \lambda_c (\partial_x\varphi )^2+\lambda_c^{-1}(\partial_y \varphi)^2\right)\right\}
\label{XY}
\end{eqnarray}
For simplicity, we have set $\mathbf{A}=0$; $\lambda_c$ and $\lambda_s$ are (finite) anisotropies.
The low energy physics is governed by two  anisotropic $XY$ models where the effective CDW stiffness is 
renormalized to $\kappa_{\rm eff}=\kappa/(1+\kappa |\textbf{Q}|^2/h)$. Hence, at low temperatures, the nematic fluctuations soften the CDW stiffness and the part of the phase diagram of the $4e$ N-SC is enhanced while that of the PDW  shrinks.  

In the limit $T_h \gg T$, Berg et al \cite{berg-2009} derived a phase diagram for the thermal melting of the PDW state. 
In this regime,  the phase transitions  are driven by the KT mechanism of unbinding  (in this case) double dislocations, vortices or half vortices bounded to single dislocations (see also Refs.\cite{radzihovsky-2008,podolsky-2009}). However,  for $T\gtrsim T_h$ not all of these processes are possible since in this regime the free energy of the dislocations of the smectic becomes finite and thus always proliferate\cite{nelson-1981}.
In this case, dislocations get an energy $E_d\sim \ln(\xi/a)$ where $a$ is the core of the defect and $\xi$ is a typical length scale controlled by the competition between the Frank constant $K_3$ and the CDW stiffness $\kappa$,  $\xi\sim \sqrt{K_3/\kappa}$. Thus, if $T_h$ is low enough, all smectic orders (PDW and CDW) are destroyed and only nematic  and the charge $4e$ N-SC are possible.
Thus, the PDW state can only exist for $T < T_h$ and, if $T_h \to 0$, the PDW is suppressed at all $T>0$.
At $T>0$ we have short ranged smectic correlations $\langle\Delta_{+\mathbf{Q}}(\mathbf{x})\Delta^*_{+\mathbf{Q}}(0)\rangle\sim \exp{(-|\mathbf{x}|/\xi)}\;\cos(2\mathbf{Q} \cdot  \mathbf{x})$,  while  the correlations of the uniform charge $4e$ N-SC exhibit quasi-long-ranged order $\langle\Delta_{4e}(\mathbf{\mathbf{x}})\Delta_{4e}^*(0)\rangle\sim 1/|\mathbf{x}|^{\eta(T)}$, with a non-universal temperature-dependent exponent $\eta(T)$. 

\paragraph{Nematic superconductor:} 
 Deep in the $4e$ N-SC phase, above the  PDW melting, the free energy becomes
\begin{equation}
F=\int d^2x\left\{ K |\mathbf{\nabla}\alpha|^2 + \rho_s|\mathbf{\nabla} \theta|^2+\frac{\rho_s N}{2} \left(\mathbf{n}\cdot \mathbf{\nabla}\theta\right)^2\right\}
\label{FNSC}
\end{equation}  
where we have set $K_3=K_1=K$. The last term in Eq.\eqref{FNSC} is due to the coupling between the SC currents and the nematic fluctuations: $N_{ij}J_iJ_j$, where $\mathbf{J}=\rho_s\mathbf{\nabla}\theta$ is the SC current. This coupling is non-polynomial in the angle $\alpha$ of the nematic order, since $\mathbf{n}=(\cos\alpha,\sin\alpha)$.  In order to minimize the free energy, the current should be locally perpendicular to the director $\mathbf{n}$. We find two types of topological configurations that minimize Eq.\eqref{FNSC}: a)  isolated disclinations, and b) (half) vortices bounded to disclinations, in such a way that $\mathbf{J}\cdot \hat n =0$ at all points. For instance the disclination $n_i=x_i/r$ and $\partial_i\theta=\epsilon_{ij} x_j/r^2$ is one such configuration. Vortices and disclinations have an {\em attractive} logarithmic  interaction, whose sign is independent of the sign of their topological charges. Therefore, once again provided $\rho_s>0$ is that is never small, the disordering of  N-SC can only be produced in two ways: a) by unbinding disclinations, which restores isotropy but does not affect the SC  (the N-SC/SC transition),  or b) by the proliferation of (half) vortices tightly bound to disclinations (the N-SC/normal transition).
The coupling to the lattice anisotropy $h$ changes  this scenario. For $h>0$ the nematic transition becomes Ising like and it is driven by the proliferation of domain walls, not by disclinations. Hence vortices are no longer bound to disclinations. This leads to the phase diagram Figs.\ref{fig1}-\ref{fig3}.
Notice that a non-SC nematic can only occur (in this regime) if $\rho_s \to 0$.

\paragraph{Quasiparticles of the $4e$ N-SC:}  
Within a Bogoliubov-deGennes approach we expect that a spin-singlet bilinear of quasiparticle Fermi fields to couple linearly to the pair field $\Delta(\textbf{r})$. In a conventional SC, BCS theory predicts that coupling to lead to a gap in the quasiparticle spectrum once the pair field acquires an expectation value. However, in the N-SC only the charge $4e$ field has an expectation value. This order parameter is equivalent to the condensation of a {\em quartet} of fermionic quasiparticles. Alternatively, we can view the charge $4e$ N-SC as a state in which a {\em composite} operator of pair fields has an expectation value, {\it i.e.} a {\em pairing of pairs}. Thus, in the N-SC the pair field remains strongly fluctuating and uncondensed. Since the pair field couples directly  to the Bogoliubov quasiparticles,
unlike a conventional (paired) SC, there is no net energy gap in the N-SC. Nevertheless, the strongly fluctuating
pair field of this phase  strong scatters the quasiparticles resulting in a reduction of their spectral density,  an  effective ``pseudogap''.

\paragraph{External currents:} 
Consider a current $\mathbf{J}$ in the $4e$ N-SC state. The coupling between nematic and SC currents is  given by a term of the free energy of the form 
$(\rho_s/2)\; {\rm Tr} (\hat N \hat J),$ where $\hat N$ is the nematic tensor and $\hat J_{ij}=J_iJ_j-\frac{1}{2} J^2\delta_{ij}$. Thus, a constant current $\langle\mathbf{J}\rangle$, acts as an  external nematic field that explicitly breaks continuous rotational symmetry. This term  competes with the potential $V(\mathbf{n})$. 
Integrating over nematic fluctuations and  considering, as before, $\hat n_0$ along the $x$ direction and expanding $\delta\mathbf{n}(\alpha)$  to  $O(\alpha(x)^2)$ we obtain the effective strength of the symmetry breaking potential, 
$h_{\rm eff}=h+\gamma_2 (J_{y}^2-J_{x}^2)$. The effect of a current along the director 
is to soften  the smectic order. In this way,  it may be possible to  tune the effective CDW stiffness $\kappa_{\rm eff}$ and the relative weight of PDW, $4e$ N-SC and nematic phases.   

Here we  discussed the phase diagram for  PDW/N-SC phase transitions as the strength of coupling to the underlying lattice is varied. 
So far the only known physical system in which a case for PDW order can be made is the cuprate  {\LBCO} near doping $1/8$, which would put it on the right hand side of Fig.\ref{fig1}. The observation of ``fluctuating stripe order'' and nematic order in underdoped cuprates\cite{fradkin-2010} leads us to suspect that a fraction of these phase diagrams of Figs.\ref{fig1}\ref{fig3}  may possibly hide  in the pseudogap regime, including a form of charge $4e$ N-SC order. These phases may be masked by the effects of disorder as ``glassy'' regimes.

We thank E. Berg and S. Kivelson for discussions. D.G.B. thanks ICMT for hospitality. 
This work was partially supported by 
CNPq and FAPERJ (Brazil),
by the NSF, under grant DMR 0758462, and by DOE, under Contract
DE-FG02-91ER45439 through the  Materials Research Laboratory of the University of Illinois.

%


\end{document}